\documentclass[english,runningheads,a4paper]{llncs}
\usepackage{graphicx}
\usepackage{subfigure}
\usepackage{url}
\usepackage{multirow}
\usepackage{float}
\usepackage{tabularx}
\usepackage{makecell}
\usepackage{tabulary}
\usepackage{hyperref}
\usepackage{hhline}
\usepackage{array}
\usepackage{amsmath}
\usepackage{diagbox}
\usepackage{color}
\usepackage{textcomp}
\usepackage{comment}
\usepackage{fancyhdr, blindtext}
\usepackage[symbol]{footmisc}

\newcolumntype{"}{@{\hskip\tabcolsep\vrule width 1pt\hskip\tabcolsep}}

\usepackage{tikz}
\def\checkmark{\tikz\fill[scale=0.4](0,.35) -- (.25,0) -- (1,.7) -- (.25,.15) -- cycle;}

\begin{document}

\title{Recognition of Instrument-Tissue Interactions in Endoscopic Videos via Action Triplets}

\titlerunning{Surgical Action Triplet Recognition}
\author{Chinedu Innocent Nwoye\inst{1} 
\and Cristians Gonzalez\inst{2}
\and Tong Yu\inst{1}
\and Pietro Mascagni\inst{1,3} 
\and Didier Mutter\inst{2} 
\and Jacques Marescaux\inst{2}
\and Nicolas Padoy\inst{1}}


\authorrunning{Nwoye C.I. et al.}
%
\urldef{\mailsa}\path{{nwoye | npadoy}@unistra.fr}
\institute{ICube, University of Strasbourg, CNRS, IHU Strasbourg, France\\ \mailsa
\and University Hospital of Strasbourg, IRCAD, IHU Strasbourg, France
\and Fondazione Policlinico Universitario Agostino Gemelli IRCCS, Rome, Italy}

\maketitle              
\begin{abstract}
Recognition of surgical activity is an essential component to develop context-aware decision support for the operating room. 
In this work, we tackle the recognition of fine-grained activities, modeled as action triplets \textlangle{}\textit{instrument, verb, target}\textrangle{} representing the tool activity. 
To this end, we introduce a new laparoscopic dataset, \textit{CholecT40}, consisting of 40 videos from the public dataset Cholec80 in which all frames have been annotated using 128 triplet classes. 
Furthermore, we present an approach to recognize these triplets directly from the video data. 
It relies on a module called {\it class activation guide}, which uses the instrument activation maps to guide the verb and target recognition. 
To model the recognition of multiple triplets in the same frame, we also propose a trainable {\it 3D interaction space}, which captures the associations between the triplet components. 
Finally, we demonstrate the significance of these contributions via several ablation studies and comparisons to baselines on CholecT40.  

\end{abstract}

\footnotetext{Accepted at International Conference on Medical Image Computing and Computer-Assisted Intervention MICCAI 2020.}
\section{Introduction}
The recognition of the surgical workflow has been identified as a key research area in surgical data science \cite{maier2017surgical}, as this recognition enables the development of intra- and post-operative context-aware decision support tools fostering both surgical safety and efficiency. 
Pioneering work in surgical workflow recognition has mostly focused on phase recognition from endoscopic video \cite{phase_lo2003episode,phase_blum2010modeling,phase_dergachyova2016automatic,twinanda_endonet_ieee2017,phase_zisimopoulos2018deepphase,phase_funke2018temporal} and from ceiling mounted cameras \cite{twinanda2015data,chakraborty2013video}, gesture recognition from robotic data (kinematic \cite{dipietro_recognizing_miccai2016,dipietro_segmenting_ijcars2019}, video \cite{zia_surgical_miccai2018,kitaguchi_real_sr2019}, system events \cite{malpani2016system}) and event recognition, such as the presence of smoke or bleeding \cite{loukas2015smoke}. 

\begin{figure}[tb]
\includegraphics[width=1.0\linewidth]{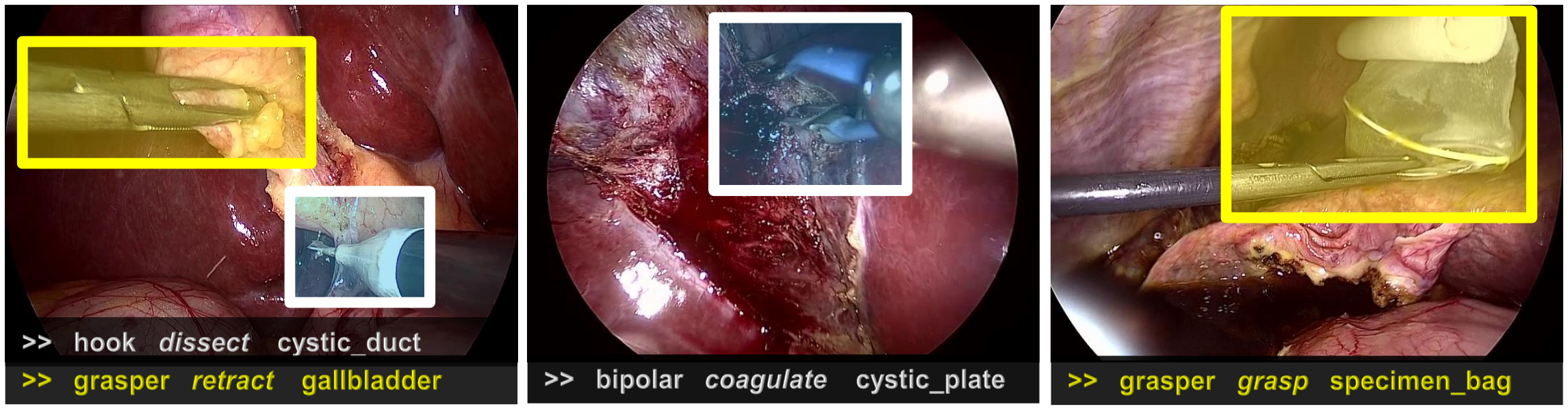} 
\caption{Examples of action triplets from the CholecT40 dataset. The three images show four different triplets. The localization is not part of the dataset, but a representation of the weakly-supervised output of our recognition model.}
\label{fig:task_and_data}
\end{figure}

In this paper, we focus on recognizing fine-grained activities representing the instrument-tissue interactions in endoscopic videos. These interactions are modeled as triplets $\langle instrument, verb, target\rangle$. Triplets represent the used instrument, the performed action, and the anatomy acted upon, as proposed in existing surgical ontologies  \cite{neumuth_acquisition_icdesa2006,katic_lapontospm_ijcars2015}. The target anatomy, while more challenging to annotate, adds substantial semantics to the recognized action/instrument. Triplet information has already been used to recognize phases \cite{katic_lapontospm_ijcars2015}, however, to the best of our knowledge, this is the first work aiming at recognizing triplets directly from the video data. The fine-grained nature of the triplets also makes this recognition task very challenging. For comparison, the action recognition task introduced within the Endovis challenge at MICCAI 2019 targeted the recognition of 4 verbs only (\textit{grasp, hold, cut, clip}).

To perform this work, we present a new dataset, called \textit{CholecT40}, containing 135K action triplets annotated on 40 cholecystectomy videos from the public Cholec80 dataset \cite{twinanda_endonet_ieee2017}. The triplets belong to 128 action triplet classes, composed of 6 instruments, 8 verbs, and 19 target classes.
Examples of such action triplets are {\textlangle{\it grasper, retract, gallbladder}\textrangle, \textlangle{\it scissor, cut, cystic duct}\textrangle, \textlangle{\it hook, coagulate, liver bed}\textrangle} (see also Fig. \ref{fig:task_and_data}).

To design our recognition model, we build a multitask learning (MTL) network with three branches for the instrument, verb and target recognition. 
We also observe that triplets are instrument-centric: an action is only performed if an instrument is present. Indeed, clinically an action can only occur if a hand is manipulating the instrument. We therefore introduce a new module, called {\it class activation guide (CAG)}, which uses the weak localization information from the instrument activation maps to guide the recognition of the verbs and targets. 
The idea is similar to \cite{gkioxari_hoi_cvpr2018}, which uses the human's ROI produced by FasterRCNN to inform the model on the likely location of the target. Other related works from the computer vision community \cite{xu_hoi_cvpr2019,qi_hoi_eccv2018,shen_hoi_wacv2018} rely heavily on the overlap of the \textit{subject-object} bounding boxes to learn the interactions. However, in addition to the fact that our work target triplets, our approach differs in that it does not rely on any spatial annotations in the dataset, which are expensive to generate.

Since instrument, verb, and target are multi-label classes, another challenge is to model their associations within the triplets. As will be shown in the experiments, naively assigning an ID to each triplet and classifying the IDs is not effective, due to the large amount of combinatorial possibilities. 
In \cite{xu_hoi_cvpr2019,qi_hoi_eccv2018,shen_hoi_wacv2018} mentioned above, \textit{human} is considered to be the only possible subject of interaction. Hence, in those works data association requires only bipartite matching to match verbs to objects. This is solvable by using the outer product of the detected object's logits and detected verb's logits to form a 2D matrix of interaction at test time \cite{shen_hoi_wacv2018}. Data association's complexity increases however with a triplet. Solving a triplet relationship is a tripartite graph matching problem, which is an NP-hard optimisation problem. In this work, inspired by \cite{shen_hoi_wacv2018}, we therefore propose a {\it 3D interaction space} to recognize the triplets.
Unlike \cite{shen_hoi_wacv2018}, where the data association is not learned, our interaction space learns the triplet relationships.

In summary, the contributions of this work are as follows:
\begin{enumerate}
    \item We propose the first approach to recognize surgical actions as triplets of \textit{\textlangle{}instrument, verb, target\textrangle{}} directly from surgical videos.
    \item We present a large endoscopic action triplet dataset, CholecT40, for this task.
    \item We develop a novel deep learning model that uses weak localization information from tool prediction to guide  verb and target detection.
    \item We introduce a trainable 3D interaction space to learn the relationships within the triplets.
\end{enumerate}

\section{Cholecystectomy Action Triplet Dataset} \label{sec:dataset}
To encourage progress towards the recognition of instrument-tissue interactions in laparoscopic surgery, we generated a dataset consisting of 40 videos from Cholec80 \cite{twinanda_endonet_ieee2017} annotated with action triplet information. We call this dataset \textit{CholecT40}. 
The cholecystectomy recordings were first annotated by a surgeon using the software \textit{Surgery Workflow Toolbox-Annotate} from the B-com institute.
For each identified \textit{action}, the surgeon sets times for the start and end frames, then labels the \textit{instrument}, the \textit{verb} and the \textit{target}. 
Any change in the triplet configuration marks the end of the current action and the beginning of a different one. 
This first step was followed by a mediation on the annotations and a class grouping carried out by another clinician.
The resulting action triplets span 128 classes encompassing 6 instruments, 8 verbs, and 19 target classes. For our experiments, we downsample the videos to 1 fps yielding a total of 83.2K frames annotated with 135K action-triplet instances. 
Table \ref{table:data_stat} shows the frequency of occurrence of the instruments, verbs and targets in the dataset. When a tool is idle, the verb and the target are both set to {\it null}.
Additional statistics on the co-occurence distribution of the triplets are presented in the supplementary material.
The video dataset is randomly split into training (25 videos, 50.6K frames, 82.4K triplets),  validation (5 videos, 10.2K frames, 15.9K triplets) and testing (10 videos, 22.5K frames, 37.1K triplets) sets.

\begin{table}[!hbtp]
\begin{center}
\begin{tabular}{lr@{\hskip 0.1in}"@{\hskip 0.1in}lr@{\hskip 0.1in}"@{\hskip 0.1in}l@{\hskip 0.1in}r@{\hskip 0.05in}|@{\hskip 0.05in}l@{\hskip 0.1in}r@{\hskip 0.05in}|@{\hskip 0.05in}l@{\hskip 0.1in}r}
\multicolumn{2}{c@{\hskip 0.1in}"@{\hskip 0.1in}}{Instrument} & \multicolumn{2}{c@{\hskip 0.1in}"@{\hskip 0.1in}}{Verb} & \multicolumn{6}{c}{Target}         \\ \hline
Name & Count & Name & Count & ID & Count & ID & Count & ID & Count
\\ \hline
grasper           & 76196   & null           & 5807   & 0 & 5807  & 8  & 236   & 16 & 88 \\
bipolar           & 5616    & place/pack     & 273    & 1 & 1169  & 9  & 137   & 17 & 114 \\
hook              & 44413   & grasp/retract  & 74720  & 2 & 75331 & 10 & 1950  & 18 & 9 \\
scissors          & 1856    & clip           & 2578   & 3 & 5173  & 11 & 5793  &    &    \\
clipper           & 2851    & dissect        & 42851  & 4 & 4378  & 12 & 8815  &    &    \\
irrigator         & 4522    & cut            & 1544   & 5 & 10023  & 13 & 641   &    &    \\
                  &         & coagulate      & 4306   & 6 & 552   & 14 & 745   &    &    \\
                  &         & clean          & 3375   & 7 & 14433 & 15 & 60    &    &      
\end{tabular}
\end{center}
\caption{Dataset statistics showing the frequency of occurrence of the instruments, verbs and targets. Target ids {\it 0...18} correspond to \textit{null, abdominal wall/cavity, gallbladder, cystic plate, cystic artery, cystic duct, cystic pedicle, liver, adhesion, clip, fluid, specimen bag, omentum, peritoneum, gut, hepatic pedicle, tissue sampling, falciform ligament, suture.}}
    \label{table:data_stat}
\end{table}

\section{Methodology} \label{sec:methods}

\begin{figure}[tb]
\includegraphics[width=0.98\linewidth]{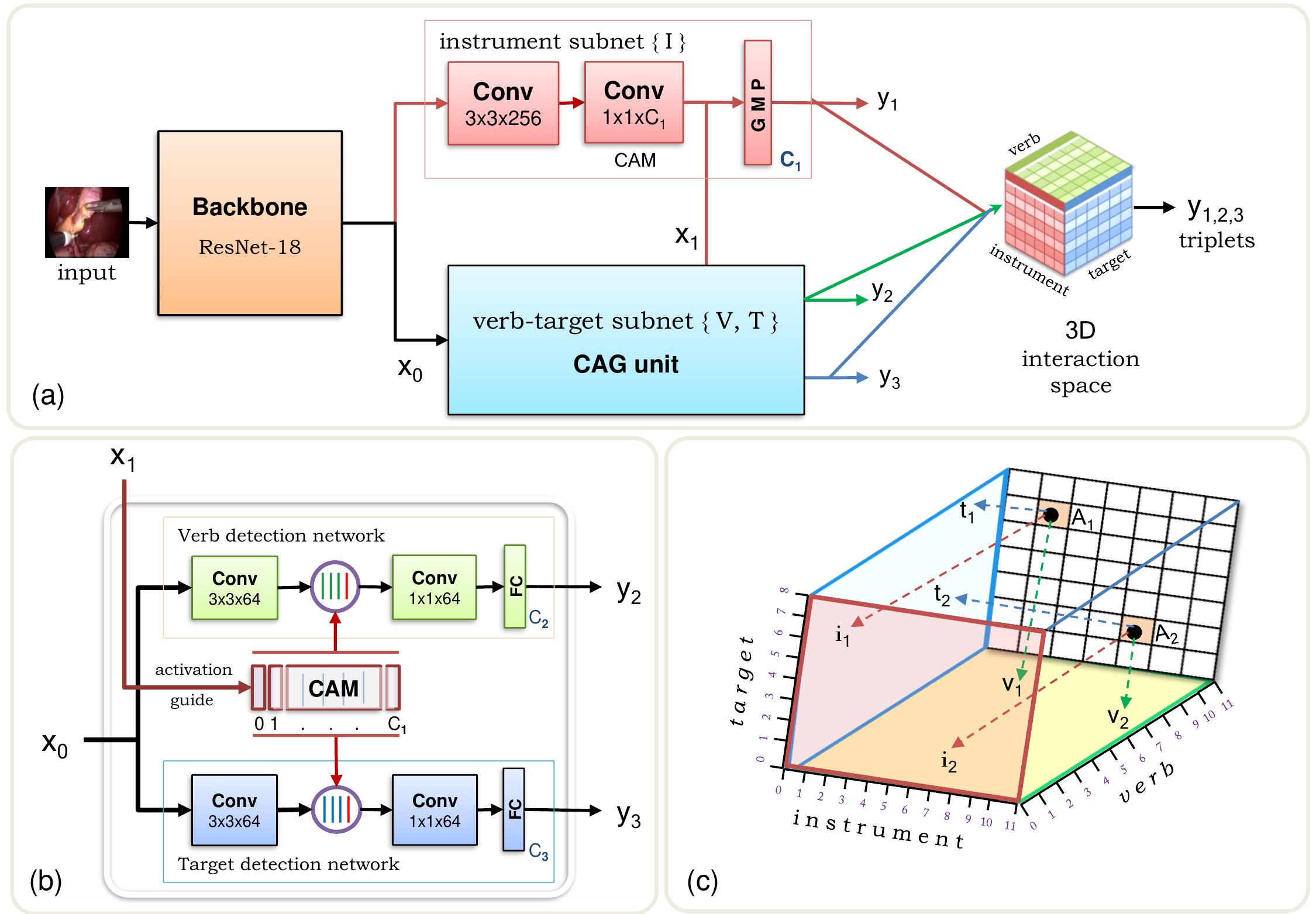}
\caption{Proposed model: (a) Tripnet for action triplet recognition, (b) class activation guide (CAG) unit for spatially guided detection, (c) 3D interaction space for triplet association.}
\label{fig:architecture}
\end{figure}

To recognize the instrument-tissue interactions in the CholecT40 dataset, we build a new deep learning model, called \textit{Tripnet}, by following a multitask learning (MTL) strategy. The principal novelty of this model is its use of the instrument's class activation guide and 3D interaction space to learn the relationships between the components of the action triplets.

\subsubsection{Multitask Learning: }
Recently, multitask deep learning models have shown that correlated tasks can share deep learning layers and features to improve performance \cite{mondal2019multitask,jin2020multi}. Following this observation, we build a MTL network with three branches for the instrument (I), verb (V), and target (T) recognition tasks. 
The instrument branch is a two layers convolutional network trained for  instrument classification. It uses global max pooling (GMP) to learn the class activation maps (CAM) of the instruments for their weak localization, as suggested in \cite{nwoye_convlstm_ijcars2019}. 
Similarly, the verb and the target branches learn the verb and target classifications using each two convolutional layers and one fully-connected (FC)-layer. 
All the three branches share the same ResNet-18 backbone for feature extraction.

\subsubsection{Class Activation Guide: }
The pose of the instruments is indicative of their interactions with the tissues. However, there is no bounding box annotation in the dataset that could be used to learn how to crop the action's locations, as done in \cite{gkioxari_hoi_cvpr2018,xu_hoi_cvpr2019,qi_hoi_eccv2018,shen_hoi_wacv2018}. We therefore hypothesize that the instrument's CAM from the instrument branch, learnable in a weakly supervised manner, has sufficient information to direct the verb and target detection branches towards the likely region of interest of the actions. 
For convenience, we regroup the three branches of the MTL into two subnets: the \textit{instrument} subnet and the \textit{verb-target} subnet, as illustrated in Fig. \ref{fig:architecture}a.
The verb-target subnet is then transformed to a {\it class activation guide (CAG)} unit, as shown in Fig. \ref{fig:architecture}b. It receives the instrument's CAM as additional input. 
This CAM input is then concatenated with the verb and target features, concurrently, to guide and condition the model search space of the verb and target on the instrument appearance cue.

\subsubsection{3D Interaction Space: }
Recognizing the correct action triplets involves associating the right $(I,V,T)$ components using the raw output vectors, also called logits, of the instrument $(I)$, verb $(V)$ and target $(T)$ branches.
In the existing work \cite{shen_hoi_wacv2018}, where the data association problem involves only the \textit{object-verb} pair, the outer product of their logits is used to form a 2D matrix of component interaction at test time.
In a similar manner, we propose a {\it 3D interaction space} for associating the triplets, as shown in Fig. \ref{fig:architecture}c. 
Unlike in \cite{shen_hoi_wacv2018}, where the data association is not learned by the trained model, we model a trainable interaction space.
Given the $m$-logits, $n$-logits and $p$-logits for the I,V,T respectively, we learn the triplets $y$ using a 3D projection function $\Psi$ as follows:
\begin{equation}
    y = \Psi(\alpha I, \beta V, \gamma T),
    \label{eqn:space}
\end{equation}
where $\alpha$, $\beta$, $\gamma$, are the learnable weight vectors for projecting I, V and T to the 3D space and $\Psi$ is an outer product operation. 
This gives an $m\times n\times p$ grid of logits with the three axes representing the three components of the triplets. For all $i \in I, v \in V, t \in T$ the 3D point $y_{i,v,t}$ represents a possible triplet. A 3D point with a probability above a threshold is considered a valid triplet.
In practice, there are more 3D points in the space than valid triplets in the CholecT40 dataset. 
Therefore, we mask out the invalid points, obtained using the training set, at both train and test times.

\subsubsection{Proposed Model: }
The proposed network is called \textit{Tripnet} and shown in Fig. \ref{fig:architecture}(a): it is an integration of the CAG unit and of the 3D interaction space within the MTL model. 
The whole model is trained end-to-end using a warm-up parameter which allows the instrument subnet to learn some semantics for a few epochs before guiding the verb-target subnet with instrument cues.

\section{Experiments}

\subsubsection{Implementation Details: }
We perform our experiments on CholecT40. 
During training, we employ three types of data augmentation (rotation, horizontal flipping and patch masking) with no image preprocessing.
The model is trained on images resized to $256\times448\times3$. 
All the individual tasks are trained for multi-label classification using the weighted sigmoid cross-entropy with logits as loss function, regularized by an $L_2$ norm with $1e^{-5}$ weight decay.
The class weights are calculated as in \cite{nwoye_convlstm_ijcars2019}.
The Resnet-18 backbone is pretrained on Imagenet. All the experimented models are trained using learning rates with exponential decay and initialized with the values $1e^{-3},1e^{-4},1e^{-5}$ for the subnets, backbone, and 3D interaction space, respectively.
The learning rates and other hyperparameters are tuned from the validation set using grid search.
Our network is implemented using TensorFlow and trained on GeForce GTX 1080 Ti GPUs.

\subsubsection{Tasks and Metrics: }
To evaluate the capacity of a model to recognize correctly a triplet and its components, we use two types of metrics:
\begin{enumerate}
    \item {Instrument detection performance:}
    This measures the average precision (AP) of detecting the correct \textit{instruments}, as the area under the precision-recall curve per instrument($AP_I$).
    \item {Triplet recognition performance:} 
    This measures the AP of recognizing the instrument-tissue interactions by looking at different sets of triplet components. We use three metrics: the \textit{instrument-verb ($AP_{IV}$), instrument-target ($AP_{IT}$)}, and \textit{instrument-verb-target ($AP_{IVT}$)} metrics. All the listed components need to be correct during the AP computation. $AP_{IVT}$ evaluate the recognition of the complete triplets. 
\end{enumerate}

\subsubsection{Baselines: }
We build two baseline models. The naive CNN baseline is a ResNet-18 backbone with two additional 3x3 convolutional layers and a fully connected (FC) classification layer with $N$ units, where $N$ corresponds to the number of triplet classes ($N=128)$. 
The naive model learns the action-triplets using their IDs without any consideration of the components that constitute the triplets. We therefore also include an MTL baseline built with the $I$, $V$ and $T$ branches described in Section \ref{sec:methods}. The outputs of the three branches are concatenated and fed to an FC-layer to learn the triplets.
For fair comparison, the two baselines share the same backbone as Tripnet.

\subsubsection{Quantitative Results: }
\begin{table}[!htbp]
    \begin{center}
    \begin{tabular}{l|@{\hskip 0.05in}c@{\hskip 0.1in}c@{\hskip 0.1in}c@{\hskip 0.1in}c@{\hskip 0.1in}c@{\hskip 0.1in}c@{\hskip 0.1in}@{\hskip 0.001in}|c@{\hskip 0.1in}}
    \multirow{2}{*}{} & \multicolumn{6}{c|}{Instrument} & \multicolumn{1}{c}{} \\\cline{2-7}
    Model & 0 & 1 & 2 & 3 & 4 & 5 & Mean \\
    \hline
    Naive CNN    & 75.3 & 04.3 & 64.6 & 02.1 & 05.5 & 06.0 & 27.5  \\
    MTL          & 96.1 & \textbf{91.9} & \textbf{97.2} & 55.7 & 30.3 & 76.8 & 74.6  \\
    \hline
    Tripnet & \textbf{96.3} & 91.6 & \textbf{97.2} & \textbf{79.9} & \textbf{90.5} & \textbf{77.9} & \textbf{89.7}
    \end{tabular}
    \end{center}
    \caption{Instrument detection performance of ($AP_{I}$) across all triplets. The IDs {0...5} correspond to {\it grasper, bipolar, hook, scissors, clipper and irrigator}, respectively.}
    \label{table:ap_i}
\end{table}

Table \ref{table:ap_i} presents the AP results for the instrument detection across all triplets. The results show that the naive model does not understand the triplet components. This comes from the fact that it is designed to learn the triplets using their IDs: two different triplets sharing the same instrument or verb still have different IDs. On the other hand, the MTL and Tripnet networks, which both model the triplet components, show competing performance on instrument detection. 
Moreover, Tripnet outperforms the MTL baseline by $15.1\%$ mean AP. This can be attributed to its use of CAG unit and 3D interaction space to learn better semantic information about the instrument behaviors.

\begin{table}[!htbp]
    \begin{center}
    \begin{tabular}{l@{\hskip 0.1in}|@{\hskip 0.05in}c@{\hskip 0.1in}c@{\hskip 0.1in}c@{\hskip 0.1in}|c@{\hskip 0.1in}}
    Model & $AP_{IV}$ & $AP_{IT}$ & $AP_{IVT}$ & Mean \\
    \Xhline{1\arrayrulewidth}
    Naive CNN    & 7.54 & 6.89 & 5.88 & 6.77 \\
    MTL          & 14.02 & 7.15 & 6.43 & 9.20 \\
    \hline
    Tripnet & \textbf{35.45} & \textbf{19.94} & \textbf{18.95} & \textbf{24.78}
    \end{tabular}
    \end{center}
    \caption{Action triplet recognition performance for instrument-verb ($AP_{IV}$), instrument-target ($AP_{IT}$) and instrument-verb-target ($AP_{IVT}$) components.}
    \label{table:results_assoc}
\end{table}

The triplet recognition performance is presented in Table \ref{table:results_assoc}. The naive CNN model has again the worst performance for the $AP_{IV}$, $AP_{IT}$ and $AP_{IVT}$ metrics, as expected from the previous results. 
The MTL baseline model, on the other hand, performs only slightly above the naive model despite its high instrument detection performance in Table \ref{table:ap_i}. 
This is because the MTL baseline model, after learning the components of the triplets, dilutes this semantic information by concatenating and feeding the output to an FC-layer. However, Tripnet improves over the MTL baseline by leveraging the instrument cue from the CAG unit. It also learns better triplet association by increasing the $AP_{IVT}$ by $12.5\%$ on average. Tripnet outperformed all the baselines in instrument-tissue interaction recognition by a minimum of $15.6\%$. In general, it can be observed that it is easier to learn the instrument-verb components than the instrument-target components. This is likely due to the fact that 
(a) a verb has a more direct association to the instrument creating the action 
(b) the dataset contains many more target classes than verb classes 
(c) many anatomical structures in the abdomen are usually discriminated with difficulty by non-medical experts.

While the action recognition performance appears to be low, it follows the same pattern as other models in the computer vision literature on action datasets of even lesser complexity. For instance, on the HICO-DET dataset \cite{chao_hoi_wacv2018}, \cite{gkioxari_hoi_cvpr2018} achieves $10.8\%$, \cite{qi_hoi_eccv2018} achieves $14.2\%$ and \cite{xu_hoi_cvpr2019} achieves $15.1\%$ action recognition AP, also known as $AP_{role}$. In fact, the current state-of-the-art performance on HICO-DET dataset is $21.2\%$ as reported on the leaderboard server. Similarly, the winner of the MICCAI 2019 subchallenge on action recognition, involving only four verb classes, scores $23.3\%$ F1-score. This shows the challenging nature of fine-grained action recognition.

\subsubsection{Ablation Studies: }
\begin{table}[!htbp]
    \begin{center}
    \vspace{-5mm}
    \begin{tabular}{l@{\hskip 0.1in}c@{\hskip 0.1in}c@{\hskip 0.1in}c@{\hskip 0.1in}@{\hskip 0.001in}|@{\hskip 0.1in}r@{\hskip 0.1in}r@{\hskip 0.1in}r@{\hskip 0.1in}r@{\hskip 0.1in}}
    FC & 3D(untrained) & 3D(trained) & CAG & $AP_{I}$ & $AP_{IV}$ & $AP_{IT}$ & $AP_{IVT}$ \\
    \Xhline{1\arrayrulewidth}
    \checkmark &  &  &  & 74.6 & 14.02 & 7.15 & 6.43 \\
     & \checkmark &  &  & 89.3 & 14.28 & 6.99 & 6.03  \\
     & \checkmark &  & \checkmark  & \textbf{89.7} & 16.72 & 7.62 & 6.32\\
     &  & \checkmark &  & 89.5 & 20.63 & 12.08  & 12.06\\ 
     &  & \checkmark & \checkmark  & \textbf{89.7} & \textbf{35.45} & \textbf{19.94} & \textbf{18.95}\\
     
    \end{tabular}
    \end{center}
    \caption{Ablation study for the CAG unit and 3D interaction space in Tripnet model.}
    \label{table:results_ablation}
    \vspace{-4mm}
\end{table}

Table \ref{table:results_ablation} presents an ablation study of the novel components of the Tripnet model. The CAG unit improves the $AP_{IV}$ and $AP_{IT}$ by approximately $2.0\%$ and $1.0\%$, respectively, justifying the need for using instrument cues in the verb and target recognition. We also observe that learning the instrument-tissue interactions is better with a trainable 3D projection than with either the untrained 3D space or with an FC-layer. This results in a large $6.0\%$ improvement of the $AP_{IVT}$. We record the best performance in all four metrics by combining the CAG unit and the trained 3D interaction space. The two units complement each other and improve the results across all metrics.

\subsubsection{Qualitative results: }
To better appreciate the performance of the proposed model in understanding instrument-tissue interactions, we overlay the predictions on several surgical images in Fig. \ref{fig:successfailures}. 
The qualitative results show that Tripnet does not only improve the performance of the baseline models, but also localizes accurately the regions of interest of the actions. 
It is observed that the majority of incorrect predictions are due to one incorrect triplet component. Instruments are usually correctly predicted and localized. As can be seen in the complete statistics provided in the supplementary material, it is however not straightforward to predict the verb/target directly from the instrument due to the multiple possible associations. More qualitative results are included in the supplementary material.

\begin{figure}[tb]
\includegraphics[width=1.0\linewidth,scale=0.99]{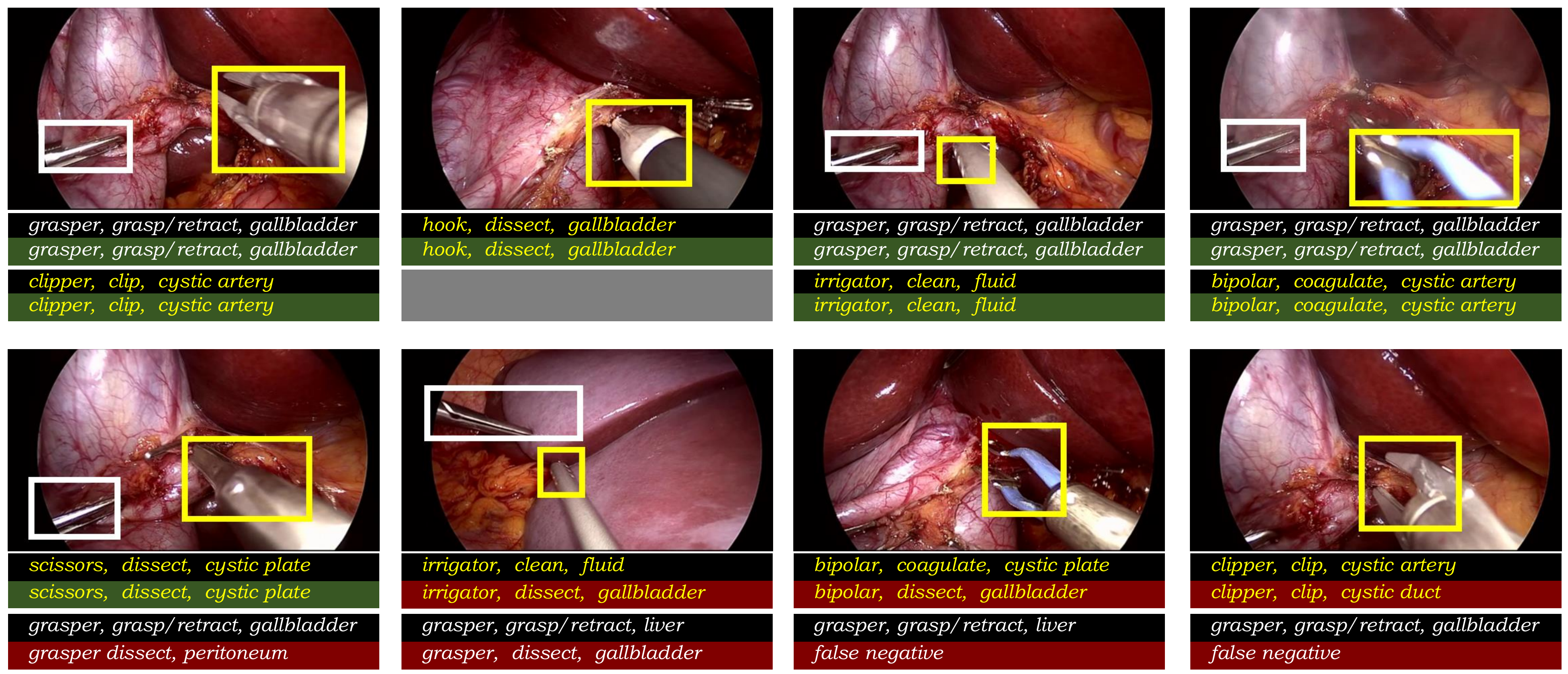}
\caption{\textbf{Qualitative results:} triplet prediction and weak localization of the regions of action (\textit{best seen in color}). Predicted and ground-truth triplets are displayed below each image: black = ground-truth, green = correct prediction, red = incorrect prediction. 
A missed triplet is marked as false negative and a false detection is marked as false positive. 
The color of the text  corresponds to the color of the associated bounding box. }
\label{fig:successfailures}
\end{figure}

\section{Conclusion} 
In this work, we tackle the task of recognizing action triplets directly from surgical videos. Our overarching goal is to detect the instruments and learn their interactions with the tissues during laparoscopic procedures. To this aim, we present a new dataset, which consists of 135k action triplets over 40 videos. For recognition, we  propose a novel model that relies on instrument class activation maps to learn the verbs and targets. We also introduce a trainable 3D interaction space for learning the \textlangle{}instrument, verb, target\textrangle{} associations within the triplets. Experiments show that our model outperforms the baselines by a substantial margin in all the metrics, hereby demonstrating the effectiveness of the proposed approach.

\subsubsection{Acknowledgements: }
This work was supported by French state funds managed within the Investissements d\textsc{\char13}Avenir program by BPI France (project CONDOR) and by the ANR (references ANR-11-LABX-0004 and ANR-16-CE33-0009).
The authors would also like to thank the IHU and IRCAD research teams for their help with the data annotation during the CONDOR project.

\bibliographystyle{splncs04}

\clearpage
\section*{\centering ===== Supplementary Material =====}
\footnotetext{Accepted at International Conference on Medical Image Computing and Computer-Assisted Intervention MICCAI 2020.}

\subsection*{\\ Appendix I : Co-occurence Distribution  of  the  Triplets}

\begin{table}[!htbp]
\begin{center}
\vspace{-4mm}
\begin{tabular}{l@{\hskip 0.1in}|@{\hskip 0.1in}r@{\hskip 0.1in}r@{\hskip 0.1in}r@{\hskip 0.1in}r@{\hskip 0.1in}r@{\hskip 0.1in}r}
\multirow{2}{*}{Verb} & \multicolumn{6}{c}{Instrument}   \\ \cline{2-7} & grasper  & bipolar  & hook & scissors & clipper  & irrigator \\ \hline
null & 2722  & 372 & 2093 &  108 &  214 &  298 \\
place/pack & 273  &   -  &   -   &  -  &   -  &   - \\
grasp/retract & 72394  & 589 & 1006 &   45  &  59 &  627 \\
clip  &    -  &   -   &  - &  - & 2578 &    - \\
dissect &  767 &  892 & 40772 &  151  &   - &  269 \\
cut  &    -  &   - &    8 & 1536  &   - &   -\\
coagulation  &   - & 3756  & 534 &   16   &  -   & - \\
clean &   40  &   7  &   -  &   -  &  - & 3328 \\
\end{tabular}
\end{center}
\caption{Dataset statistics showing the instrument-verb occurrence frequency.}
\label{table:data_stat1}
\vspace{-4mm}
\end{table}

\begin{table}[!htbp]
\begin{center}
\vspace{-4mm}
\begin{tabular}{l@{\hskip 0.1in}|@{\hskip 0.1in}r@{\hskip 0.1in}r@{\hskip 0.1in}r@{\hskip 0.1in}r@{\hskip 0.1in}r@{\hskip 0.1in}r}
\multirow{2}{*}{Target} & \multicolumn{6}{c}{Instrument}   \\ \cline{2-7} & grasper  & bipolar  & hook & scissors & clipper  & irrigator \\ \hline
null  & 2722   &  372  &  2093  &   108   &  214   &  298\\
abdominal wall/cavity  & 36  &   361    &  -   &  -   &   -  &   772\\
gallbladder  & 48720  &   731  & 25750   &   57   &   -  &   73\\
cystic plate  & 1451   &  478  &  2959   &   32   &   54  &   199\\
cystic artery  &  38   &  190  &  2639  &   558  &   953   &    -\\
cystic duct  &  786  &   215  &  6710  &   670  &  1572   &   70\\
cystic pedicle  &  112   &   90   &   48    &   4    &  58   &  240\\
liver  &  10919  &  2399   &  356 &     90    &  -  &   669\\
adhesion  &   1  &    73    &   9  &   154    &  -    &  -\\
clip  &   137   &    -  &    -    &  -   &    -   &    -\\
fluid  &    7    &  -    &  -   &   -   &    -   & 1943\\
specimen bag  &  5685   &   79   &    -    &   -  &    -  &    29\\
omentum  &  4413  &   521  &  3553  &   110   &    -  &   218\\
peritoneum  &  298   &   -  &   286   &   57   &    -  &    -\\
gut  & 709  &    19   &    6    &   -    &  -   &   11\\
hepatic pedicle  &  10   &   46  &     4    &  -   &    -   &    -\\
tissue sampling  &   72    &   9    &   -    &   7    &   -    &  -\\
fallciform ligament  &  81    &  33   &    -   &   -   &   -   &    -\\
suture  &   1   &    -  &    -   &    9   &    -   &    -\\
\end{tabular}
\end{center}
\caption{Dataset statistics showing the instrument-target occurrence frequency.}
\label{table:data_stat2}
\end{table}

\clearpage

\subsection*{Appendix II : Qualitative Results}

\begin{figure}[!htbp]
\centering
    \includegraphics[width=0.97\linewidth,scale=0.99]{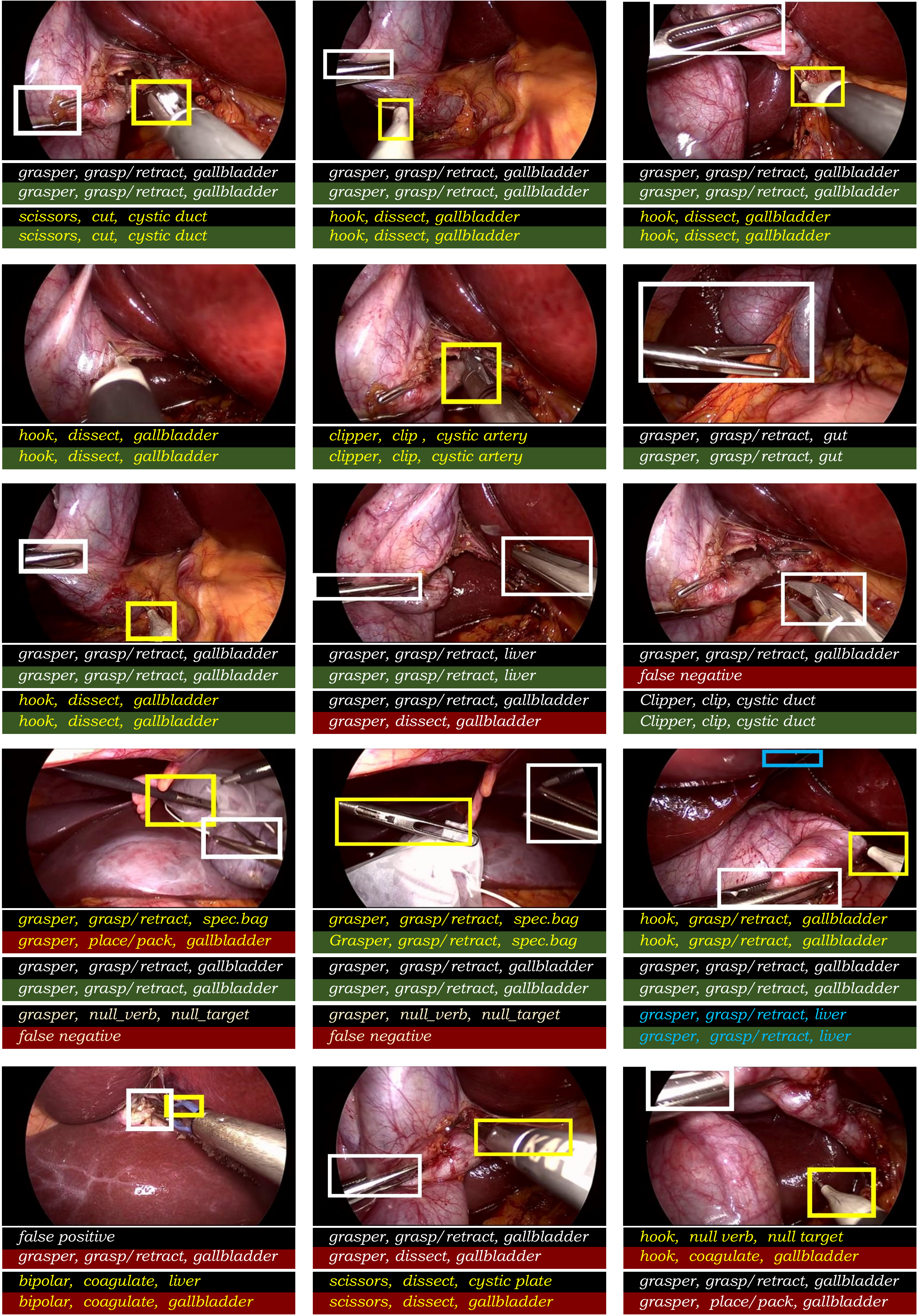}
    \caption{Additional qualitative results showing both success and failure cases.}
\label{fig:successfailures}
\end{figure}

\end{document}